\documentclass[useAMS,usenatbib]{mn2e}
\usepackage{graphicx}

\usepackage{ulem}
\title{Gamma-ray sources like V407 Cygni in Symbiotic Stars}
\author[ L\"{u} et al. ]{Guoliang L\"{u}$^{1, 2}$\thanks{E-mail:
guolianglv@gmail.com (LGL)}, Chunhua Zhu$^{1, 2, 3}$, Zhaojun Wang$^{1, 2}$, Wensheng Huo$^{1, 2}$, Yuangui Yang$^4$\\
$^{1}$School of Physics, Xinjiang University, Urumqi, 830046,
China.\\
$^2$Xinjiang University-National Astronomical Observatories Joint
Center for Astrophysics, Urumqi, 830046, China.\\
$^{3}$National Astronomical Observatories / Urumqi Observatory, the
Chinese Academy of Sciences, Urumqi, 830011, China.\\$^{4}$School of
Physics and Electronic Information Huaibei Normal University 235000
Huaibei City, Anhui Province, China.}
\begin{document}

\date{}

\pagerange{\pageref{firstpage}--\pageref{lastpage}} \pubyear{2010}

\maketitle

\label{firstpage}

\begin{abstract}
Using a simple accelerating model and an assumption that
$\gamma$-rays originate from $p-p$ collisions for a $\pi^0$ model,
we investigate $\gamma$-ray sources like V407 Cygni in symbiotic
stars. The upper limit of their occurrence rate in the Galaxy is
between 0.5 and 5 yr$^{-1}$, indicating that they may be an
important source of the high-energy $\gamma$-rays. The maximum
energies of the accelerated protons mainly distribute around
$10^{11}$ eV, and barely reach $10^{15}$ eV.  The novae occurring in
D-type SSs with ONe WDs and long orbital periods are good candidates
for $\gamma$-ray sources. Due to a short orbital period which
results in a short acceleration duration, the nova occurring in
symbiotic star RS Oph can not produce the $\gamma$-ray emission like
that in V407 Cygni.
\end{abstract}

\section{Introduction}
Symbiotic stars (SSs) are usually interacting binaries, composed of
a cool star, a hot component and a nebula. The hot component is
usually a white dwarf (WD). The cool component is either a normal
red giant (RG) in S-type or a Mira variable surrounded by an
optically thick dust shell in D-type. Symbiotic novae are a small
subclass of thermonuclear novae which occur on a WD surface fueled
by mass accreted from an RG.  They can produce super-soft or soft
X-ray emission in SSs \citep{Murset1997,Zhu2010}.

V407 Cygni is a D-type SS consisting of a Mira-type pulsating RG.
Recently, \cite{Abdo2010} reported the $Fermi$ Large Area Telescope
detection of variable $\gamma$-ray emission (0.1---10 GeV) from the
nova of SS V407 Cygni. They explained that the $\gamma$-ray spectrum
originate from proton-proton ($p-p$) interaction by $\pi^0$
model\footnote{$\pi^0$ model is that secondary neutral pion decays
$\gamma$-rays from proton-proton ($p$-$p$) collisions, i.e., the
reaction considered is $p+p\to\pi^0+X$ and the decay is $\pi^0\to
2\gamma$, where $p$ represents a proton,  $\pi^0$ represents a
neutral pion, and X represents any combination of particles (see the
references of \citet{Kamae2006})}, but inverse Compton scattering
(See \cite{Blumenthal1970} for an exclusive review) of infrared
photons from the RG by electrons cannot be ruled out. The above two
mechanisms producing $\gamma$-rays need the high-energy protons or
electrons. Cosmic rays with high energies are thought to originate
from supernovae remnants (SNRs). In order to having an efficient
acceleration mechanism the theory of diffusive shock acceleration
has been developed  (see a recent review from \cite{Malkov2001}).
Diffusive shock acceleration applies only for particles with a
Larmor radius larger than the typical shock thickness. If electrons
and protons are equilibrium in the shock, the Larmor radius of
electron is a factor $(m_{\rm e}/m_{\rm p})^{1/2}$ smaller than that
of proton, where $m_{\rm e}$ and $m_{\rm p}$ are masses of an
electron and a proton, respectively. Only electrons which are
already relativistic can cross the shock and start acceleration.
Therefore, protons are accelerated more easily than electrons under
the mechanism diffusive shock acceleration.

In this Letter, we investigate the likely possibility of the
symbiotic novae producing  $\gamma$-rays, then assume that
$\gamma$-rays originate from $p-p$ collisions for a $\pi^0$ model,
and investigate the $\gamma$-ray sources in SSs.
\section{Acceleration Model}
\label{sec:model} In general, the theory of diffusive shock
acceleration is used for SNRs to explain the emission of cosmic rays
with high energy. In this work the acceleration model used for the
symbiotic novae like V407 Cygni is similar with that used for SNRs.
There is usually a low particle-density circumstellar medium (CSM)
around SNRs. However, a symbiotic nova like V407 Cygni  is usually
embedded in a dense CSM which is mainly formed from stellar winds
lost by the red giant. This difference makes  it possible to
efficiently accelerate protons in the symbiotic novae.

According to the theory of diffusive shock acceleration for SNRs,
the maximum attainable energy for cosmic rays is determined by the
size of accelerator, the magnetic field of the CSM and the energy
losses resulting from adiabatic processes and synchrotron processes.
The size depends on the explosion evolution of SNR. According to
\cite{Kirk1994}, the explosion includes three phases: free-expansion
phase, Sedov-Taylor phase and snow-plough phase. During the
free-expansion phase, the kinetic energy of ejecta remains untapped,
and the particle acceleration is not significant. Once the mass
swept-up by the shock becomes comparable to the mass of the ejecta
$M_{\rm ejc}$, the explosion enters the Sedov-Taylor phase. The
acceleration efficiency is the highest in this phase. The maximum
energy for protons is approximately given by \citep{Schure2010}
\begin{equation}
E^{\rm p}_{\rm max}=\frac{3ZeBV_{\rm sh}^2t_{\rm ST}}{\xi_{\sigma}
c}=\frac{3ZeBV_{\rm sh}R_{\rm ST}}{\xi_{\sigma} c} \label{eq:epmax}
\end{equation}
where $Z$ is the charge number, $e$ is the elementary electric
charge, $c$ is the speed of light and $\xi_{\sigma}$ is a relation
between the compression ratio of the density and magnetic field.
Here, we assume that the magnetic field is parallel to the shock
normal, which means $\xi_{\sigma}=20$. $B$ is the magnetic field
strength, and $t_{\rm ST}\sim R_{\rm ST}/V_{\rm sh}$ is the duration
for which the particles stay in the Sedov-Taylor phase, where
$V_{\rm sh}$ is the shock speed and $R_{\rm ST}$ is the radius where
the shock sweeps up the matter whose mass is equal to that of the
ejecta.

In general, for SNRs,  the magnetic field $B$ of the CSM is $\sim
\mu$Gauss, and $R_{\rm ST}$ is $\sim$ pc due to the high mass of the
ejecta ($\sim M_\odot$) and the low particle density of CSM
\citep{McKee1995}. Typical $t_{\rm ST}$ is several hundred years.
However, in symbiotic novae like V407 Cygni, there are some physical
conditions which are greatly different from those in SNRs:
\\
(i) The magnetic field of the CSM is the magnetic field of the
stellar wind from the RG, and it is given by \cite{Bode1985}
\begin{equation}
B=\sqrt{8\pi\rho k T_{\rm g}/\bar{m}}
\end{equation}
where $k$ is Boltzmann's constant, $\bar{m} = 10^{-24}$ g is the
mean particle mass, and $T_{\rm g}$ is the temperature of the
stellar wind. V407 Cygni is a D-type SS in which the RG has dust
shells. The temperature for the dust condensation zone is $\sim$
1000 K \cite{Gail1999}. Here, we take $T_{\rm g}=1000$K. The density
$\rho$ is given by
\begin{equation}
\rho=\frac{\dot{M}_{\rm L}}{4\pi V_{\rm w}(R^2+a^2-2Ra\cos\theta)}
\end{equation}
where $\dot{M}_{\rm L}$ is the mass-loss rate of the RG, $V_{\rm w}$
is the stellar wind velocity, $a$ is the binary separation, a
distance $R$ and polar angle $\theta$ is from the WD center towards
the RG. For simplicity, we only consider $\theta=0^{\rm o}$. As
shown by \cite{Bode1985}, $B$ is $\sim 10^{-2}$  Gauss, which is
$10^{3}$ times higher than that in the CSM of SNR. \\
(ii) For a typical nova, the mass of ejecta is $\sim 10^{-6}
M_\odot$ \citep{Yaron2005}, which is much less than that of the
ejecta in a typical supernova. Furthermore, we note that the
duration of  matter ejecting in a typical nova is $\sim$ several
days or tens of days, and the matter ejected has an average
expansion velocity $V_{\rm av}$ over the whole ejecting matter phase
and a maximal expansion velocity $V_{\rm max}$. This means that a
part of the matter ejected has the high expansion velocity $V_{\rm
max}$. The shock in a nova is mainly produced by the matter ejected
with high velocity. We assume that $V_{\rm sh}=V_{\rm max}$, and use
a parameter $\eta$ to define a ratio of the mass ejected with a
velocity of $V_{\rm max}$ to the whole ejecta. $R_{\rm ST}$ can be
given by the following equation:
\begin{equation}
\eta M_{\rm ejc}=\int_{R_{\rm WD}}^{R_{\rm ST}}4\pi R^2\rho {\rm d}
R
\end{equation}
where $R_{\rm WD}$ is the radius of WD. \cite{Abdo2010} found that
the peak flux in $\gamma$-rays was observed after 3-4 days of a nova
outburst from V407 Cygni on 10 March 2010. This implies that $t_{\rm
ST}$ in V407 Cygni should shorter than 3 days. In our model $t_{\rm
ST}$ depends on the parameter $\eta$. We find that $t_{\rm ST} \sim$
days when $\eta \sim$ 0.01.

The novae occurring in SSs are surrounded by the dense stellar winds
from the RGs. They offers an environment for high efficient particle
acceleration. Therefore,  they may be an important source of the
high-energy $\gamma$-rays in the Galaxy.
\section{Symbiotic Stars}
In general, SSs are the detached interacting binaries in which the
WDs accrete the matter of the RGs via stellar winds. By a population
synthesis method, \cite{Lu2006} carried out a detailed investigation
of SSs. They found that the occurrence rate of the novae in SSs is
greatly affected by common-envelope evolution and the stellar wind
velocity $V_{\rm w}$ of the RG. Following \cite{Lu2006} and
\cite{Zhu2010}, for common-envelope evolution in different
simulations we use an $\alpha_{\rm ce}\lambda_{\rm ce}=0.5$ in
$\alpha$-algorithm and $\gamma=1.75$ in a $\gamma$-algorithm,
respectively; for the stellar wind, $V_{\rm w}=\frac{1}{2}v_{\rm
esc}$ where $v_{\rm esc}$ is the escape velocity and $V_{\rm w}$ is
determined by the relation between the mass-loss rates and the
terminal wind velocities fitted by \cite{Winters2003} as:
\begin{equation}
\log_{10} (\dot{M}/M_\odot{\rm yr}^{-1})=-7.40+\frac{4}{3}\log_{10}
(V_{\rm w}/{\rm km \, s^{-1}}). \label{eq:winters}
\end{equation}
In this work we consider three cases with different input
parameters:
\\(i) \ \ in case 1, $\alpha_{\rm ce}\lambda_{\rm ce}=0.5$ and $V_{\rm
w}=\frac{1}{2}v_{\rm esc}$;\\(ii) \ in case 2, $\gamma=1.75$ and
$V_{\rm w}=\frac{1}{2}v_{\rm esc}$;\\(iii) in case 3, $\alpha_{\rm
ce}\lambda_{\rm ce}=0.5$ and $V_{\rm w}$ taken as {\rm Eq.
(\ref{eq:winters})}.

Using the model of SSs and grid for novae in \cite{Yaron2005}, we
can estimate $E^{\rm p}_{\rm max}$ in which $\theta=0^{\rm o}$ and
$\eta=0.01$ for every nova. According to \cite{Kamae2006}, $p-p$
interaction can occur when the energy of proton is higher than
$10^9$ eV, which results in $\gamma$-ray emission. Therefore, we
assume that the novae in SSs are $\gamma$-ray sources if the $E^{\rm
p}_{\rm max}$ in Eq. (\ref{eq:epmax}) is higher than $10^9$ eV.

\begin{figure}
\includegraphics[totalheight=3.0in,width=3.0in,angle=-90]{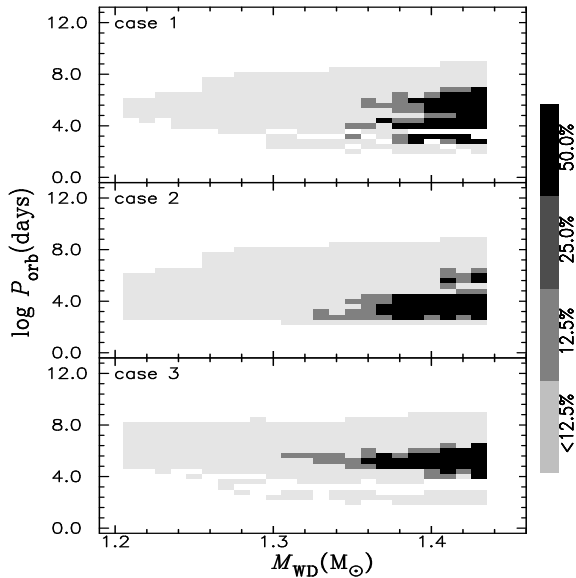}
\caption{Gray-scale maps of WD's masses vs. orbital
          periods for SSs as $\gamma$-ray sources  in cases 1, 2 and 3. The gradations of gray-scale
            correspond to the regions where the number density of systems is,
            respectively,  within 1 -- 1/2,
            1/2 -- 1/4, 1/4 -- 1/8, 1/8 -- 0 of the maximum of
             ${{{\partial^2{N}}\over{\partial {\log P_{\rm orb}}}{\partial {M_{\rm WD}}}}}$,
             and blank regions do not contain any stars.    } \label{fig:mp}
\end{figure}

\begin{figure}
\includegraphics[totalheight=3.0in,width=3.0in,angle=-90]{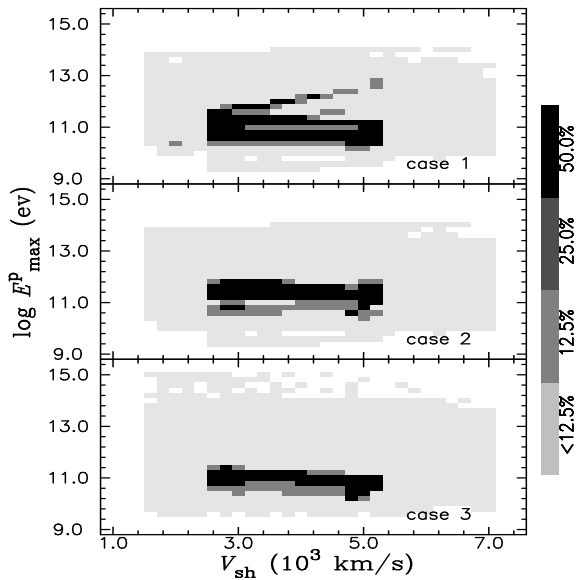}
\caption{Similar to Figure \ref{fig:mp}, but for the maximum energy
$E^{\rm p}_{\rm max}$ of the protons accelerated vs. the shock
velocity $V_{\rm sh}$. } \label{fig:egvs}
\end{figure}

\begin{figure}
\includegraphics[totalheight=3.0in,width=3.0in,angle=-90]{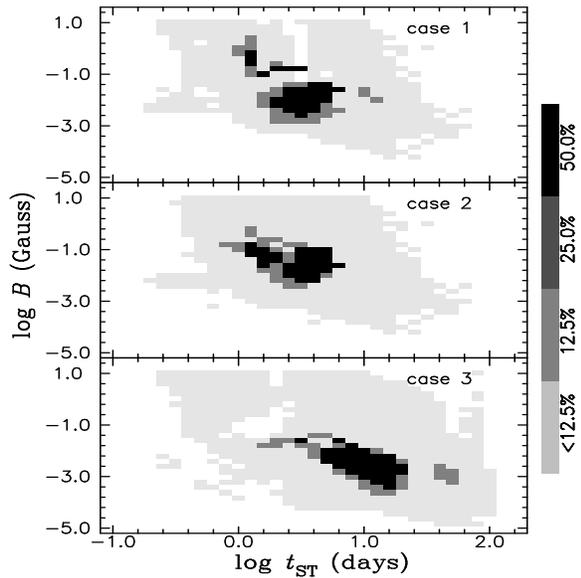}
\caption{ Similar to Figure \ref{fig:mp}, but for the duration of
Sedov-Taylor phase $t_{\rm ST}$ vs. the magnetic field of the
stellar wind from RGs. } \label{fig:tdmb}
\end{figure}
\section{Results }
Using a population synthesis method described in \cite{Lu2006,
Lu2008}, we model $10^6$ binary systems which gives a statistical
error for our Monte Carlo simulation lower than 5 percent for the
symbiotic novae. In order to estimate the occurrence rate of the
$\gamma$-ray sources like V407 Cygni, we assume one binary with
primary mass more massive than 0.8 $M_\odot$ is formed annually in
the Galaxy. We do not consider the energy losses of the accelerated
protons via adiabatic process and synchrotron process. Therefore, we
overestimate the maximum energy of the proton and the occurrence
rate of the $\gamma$-ray sources in this work.

We select symbiotic novae as $\gamma$-ray sources if $E^{\rm p}_{\rm
max}$ of the accelerated protons is larger than $10^9$ eV. Our model
shows that the upper limits of the occurrence rates of $\gamma$-ray
sources like V407 Cygni in SSs are 0.5 yr$^{-1}$ in case 1, 2.0
yr$^{-1}$ in case 2 and 5.0 yr$^{-1}$ in case 3, respectively.
Compared with the results in \cite{Lu2006}, about 15\% of the novae
in SSs for cases 1 and 2 can produce $\gamma$-ray emission, and it
is 40\% in case 3 because the wind velocity in Eq.(\ref{eq:winters})
is favorable for a strong nuclear outburst. If the Galactic cosmic
rays originate from the supernova whose occurrence in the Galaxy is
$\sim$ 0.01 yr$^{-1}$, we suggest that symbiotic novae like V407
Cygni may be another important source of the high-energy
$\gamma$-rays. However, the contribution of the symbiotic novae to
total cosmic rays cannot be known until the spectra-energy
distribution of the $\gamma$-rays from the novae is calculated,
which will be carried out in further work.

Figure \ref{fig:mp} shows the distribution of the WD's masses vs.
orbital periods for SSs as $\gamma$-ray sources. Majority of WDs in
these SSs are ONe WDs and they have masses larger than 1.3
$M_\odot$. As \cite{Lu2008} mentioned, the most significant property
of novae occurring on the surface of ONe WDs is high neon abundance
in the ejected materials. V407 Cygni may offer a chance to
investigate the thermonuclear outbursts on the surface of ONe WDs.
However, to our knowledge, there is no observational data on the
neon abundance of the ejecta in this nova from V407 Cygni. The peak
of orbital-period distribution is around $\sim 10^5 $ days in cases
1 and 3, and it is around $\sim 10^3$ days in case 2.
\cite{Munari1990} suggested that the V407 Cygni has an orbital
period of 43 years, which is consistent with our results. Novae in
SSs with long orbital periods mean that RGs have high mass-loss
rates. In our work, the majority of the RGs have mass-loss rates
higher than $5\times 10^{-7} M_\odot$ yr$^{-1}$.
\cite{Ferrarotti2006} suggested that RGs produce significant dust
when their mass-loss rates are higher than $3\times 10^{-7} M_\odot$
yr$^{-1}$. Therefore, we consider that most of SSs as $\gamma$-ray
sources have massive WDs, long orbital periods and RGs with dust
shells. This means that the novae occurring in D-type SSs with ONe
WDs are good candidates for $\gamma$-ray sources.

RS Oph is a symbiotic recurrent nova which had previously undergone
recorded outbursts in 1898, 1933, 1958, 1967 and 1985. It comprises
a red giant star in a $455.72\pm0.83$ day orbital period with a
white dwarf (WD) of mass near the Chandrasekhar limit. Recently, RS
Oph was observed to be undergoing an outburst On 2006 Februry 12.83
UT \citep{Hirosowa2006}. During the first 3 days a hard X-ray
emission (14-25 keV) was clearly detected, and there was a weak
detection in the 25-50 keV band immediately following the outburst
\citep{Bode2006}. After that, the X-ray spectrum in RS Oph was seen
to evolve from the relatively hard to a super-soft source state
(Evolutionary details are in \cite{Nelson2008}). However, there is
no conclusive evidence for gamma-ray emission like V407 Cygni in the
2006 outburst. The orbital period of V407 Cygni is $\sim$ 40 times
that of RS Oph, which means that the density of CSM around the WD in
the later is higher than 100 times of that in the former if
$\frac{\dot{M}_{\rm L}}{V_{\rm w}}$ is comparable in the two
binaries\footnote{\cite{Munari1990} estimated that the mass-loss
rate of the RG in V407 Cygni is $\sim 6\times 10^{-7} M_\odot$
yr$^{-1}$. Considering no thick dust shells around RS Oph, we assume
that the mass-loss rate of the RG in RS Oph is $\sim 10^{-7}
M_\odot$ yr$^{-1}$. However, due to the dust-driven wind
\citep{Gail1986}, the stellar wind velocity $V_{\rm w}$ in V407
Cygni is higher than that in RS Oph}. Therefore, the duration of the
diffusive shock acceleration in 2006 outburst of RS Oph, $t_{\rm
ST}$, is too short so that protons and electrons can not be
accelerated enough energy to produce $\gamma$-ray emission.

Figure \ref{fig:egvs} gives the distribution of $E^{\rm p}_{\rm
max}$ vs. $V_{\rm sh}$. The peak of $E^{\rm p}_{\rm max}$ is at
$\sim 10^{11}$ eV, and $E^{\rm p}_{\rm max}$ hardly reaches
$3\times10^{15}$ eV which is the knee of the cosmic ray spectra. If
the $\gamma$-ray energy originating from $p-p$ interaction is
comparable to $E^{\rm p}_{\rm max}$, they should be in the
low-frequency part of high-energy cosmic rays. These nuclear
outbursts are very strong so that the ejecta have high velocity. As
Figure \ref{fig:egvs} shows, $V_{\rm sh}$ in the symbiotic novae is
$\sim$ several $10^3$ km/s. According to the calculations of
\cite{Yaron2005}, in these strong nuclear outbursts most of the
accreted mater is expelled and in some cases even an erosion of the
WD occurs. Therefore, the massive WDs in $\gamma$-ray sources do not
explode as supernovae.

Figure \ref{fig:tdmb} shows the distribution of the duration of
Sedov-Taylor phase $t_{\rm ST}$ vs. the magnetic field of the
stellar wind from RGs. The peaks of the magnetic field distribution
are around $\sim 10^{-2}$ Gauss in cases 1 and 2, and it is around
$\sim 10^{-3}$ Gauss in case 3 because a high mass-loss rate results
in high $V_{\rm w}$ (See Eq. (\ref{eq:winters})). The magnetic
fields of the stellar winds around the novae in SSs are $10^3$ times
higher than those of CSM around SNRs. The peak of $t_{\rm ST}$ is at
$\sim 3$ days in cases 1 and 2, while it is around $\sim 10$ days in
case 3 due to a high $V_{\rm w}$ which results in the low density of
stellar wind. However, as mentioned in the \S \ref{sec:model},
$t_{\rm ST}$ are greatly affected by an uncertain parameter $\eta$.

\section{ Conclusions}
In this Letter, we use a toy model to investigate the $\gamma$-ray
sources which originate from $p-p$ collisions in the novae of SSs.
The symbiotic novae occurring on the surface of accreting WDs are
surrounded by the dense stellar winds from the RGs. They offers an
environment for highly efficient particle acceleration. We estimate
that the upper limit of the occurrence rate of $\gamma$-ray sources
in SSs in the Galaxy is between 0.5 and 5 yr$^{-1}$. Therefore, they
may be an important source of the high-energy $\gamma$-rays. The
maximum energies of the accelerated protons  mainly distribute
around $10^{11}$ eV, and barely reaches $10^{15}$ eV. If the
$\gamma$-ray energy originating from $p-p$ interaction is comparable
to $E^{\rm p}_{\rm max}$, they should be in the low-frequency part
of high-energy cosmic rays. In SSs as $\gamma$-ray sources, majority
of WDs are ONe WDs and have masses larger than 1.3$M_\odot$, and
most of RGs have dust shells. The novae occurring in D-type SSs with
ONe WDs are good candidates as $\gamma$-ray sources. Due to the
short orbital period, the nova occurring in RS Oph hardly produces
$\gamma$-ray emission like that in V407 Cygni.

\section*{Acknowledgments}
We thank an anonymous referee for his/her comments which helped to
improve the paper. GL thanks Dr Jamie Leech for correcting the
English language of the manuscript. This work was supported by the
National Natural Science Foundation of China under Nos. 10763001,
10963003 and 11063002, Natural Science Foundation of Xinjiang under
Nos.2009211B01 and 2010211B05, Foundation of Huoyingdong under No.
121107 and Scientific Research Program of the Higher Education
Institution of Xinjiang under No. XJEDU2008S12.


\appendix

\label{lastpage}

\end{document}